\documentclass[comment, prl,twocolumn,nofootinbib, preprintnumbers, superscriptaddress]{revtex4}
\usepackage{amsmath,amssymb,bm,slashed,braket}
\usepackage{graphicx}
\usepackage{epstopdf}
\usepackage{float}
\usepackage{xcolor}
\usepackage{multirow}
\usepackage{dsfont}
\usepackage[colorlinks=true,
            linkcolor=blue,
            urlcolor=blue,
            citecolor=purple,          
            bookmarks=true,
            bookmarksnumbered=true,
            breaklinks=true,
            pdfpagemode=Fullscreen,
            pdfstartview=FitBH]{hyperref}
\usepackage[normalem]{ulem}
\allowdisplaybreaks[4]
\usepackage[capitalise]{cleveref}
\usepackage{tikz}
\usetikzlibrary{decorations.markings}
\usepackage{orcidlink}
\usepackage{tikz} 
\usepackage{tkz-euclide}
\usetikzlibrary{backgrounds} 
\usetikzlibrary{decorations.pathmorphing}
\usetikzlibrary{arrows.meta}
\tikzset{
mystyle/.style={line width=1, baseline, scale=0.6, every node/.style={scale=1}},
v/.style={decorate, draw, decoration={snake, segment length=2.mm, amplitude=0.5mm}},
f/.style={draw, decoration={markings,mark=at position #1 with {\arrow[]{Latex[length=1.5mm,width=1.5mm]}}},
    postaction={decorate},node contents=#1},
f/.default=.6,
fb/.style={draw,decoration={markings,mark=at position #1 with {\arrowreversed[]{Latex[length=1.5mm,width=1.5mm]}}},
    postaction={decorate},node contents=#1},
fb/.default=.6,
s/.style={dashed,draw, decoration={markings,mark=at position #1 with {\arrow[]{Latex[length=1.5mm,width=1.5mm]}}},
    postaction={decorate},node contents=#1},
s/.default=.6,    
sb/.style={dashed,draw,decoration={markings,mark=at position #1 with {\arrowreversed[]{Latex[length=1.5mm,width=1.5mm]}}},
    postaction={decorate},node contents=#1},
sb/.default=.4,
snar/.style={dashed,draw,line width =1.25pt},
cross/.style={cross out, draw=black, minimum size=2*(#1-\pgflinewidth), inner sep=0pt, outer sep=0pt}, 
         }

\newcommand{\C}{ {\tt C} }
\newcommand{\tL}{ {\tt L} }
\newcommand{\tR}{ {\tt R} }
\newcommand{\hc}{\text{H.c.}}

\begin{document}

\title{New chiral structures for baryon number violating nucleon decays}
\author{Yi Liao\,\orcidlink{0000-0002-1009-5483}}
\email{liaoy@m.scnu.edu.cn}
\author{Xiao-Dong Ma\,\orcidlink{0000-0001-7207-7793}}
\email{maxid@scnu.edu.cn}
\author{Hao-Lin Wang\,\orcidlink{0000-0002-2803-5657}}
\email{whaolin@m.scnu.edu.cn}
\affiliation{State Key Laboratory of Nuclear Physics and
Technology, Institute of Quantum Matter, South China Normal
University, Guangzhou 510006, China}
\affiliation{Guangdong Basic Research Center of Excellence for
Structure and Fundamental Interactions of Matter, Guangdong
Provincial Key Laboratory of Nuclear Science, Guangzhou
510006, China}

\begin{abstract}
We examine the most general nucleon decay interactions that involve three light quarks without being acted upon by a derivative. 
We identify four generic operator structures that correspond to the irreducible representations in the chiral group ${\rm SU(3)}_{\tt L}\otimes {\rm SU(3)}_{\tt R}$ of QCD, 
\{$\pmb{8}_{\tt L}\otimes \pmb{1}_{\tt R}$,
$\bar{\pmb{3}}_{\tt L}\otimes \pmb{3}_{\tt R}$,
$\pmb{6}_{\tt L}\otimes \pmb{3}_{\tt R}$,
$\pmb{10}_{\tt L}\otimes \pmb{1}_{\tt R}$\}, plus their chirality partners under the interchange of chiralities ${\tt L}\leftrightarrow {\tt R}$. 
While half of them have been extensively discussed in the literature, the other half, $\pmb{6}_{\tt L(R)}\otimes \pmb{3}_{\tt R(L)}$ and
$\bar{\pmb{10}}_{\tt L(R)}\otimes \pmb{1}_{\tt R(L)}$, are identified for the first time. 
We perform chiral matching for these interactions at the leading chiral order and find that each has a unique chiral realization in terms of the octet baryons and pseudoscalars. Notably, the chiral interaction in the $\pmb{6}_{\tt L(R)}\otimes \pmb{3}_{\tt R(L)}$ representation appears at the same chiral order as those of the known ones, while the one in the $\pmb{10}_{\tt L(R)}\otimes \pmb{1}_{\tt R(L)}$ representation appears at a higher chiral order. These new structures are prevalent in effective field theories and ultraviolet models, and they offer novel experimental avenues to search for baryon number violating nucleon decays. 
\end{abstract}

\maketitle 

\vspace{0.1cm}{\bf Introduction.}
The stability of matter and the very existence of ourselves imply that the baryon number must be an almost conserved quantity --- while the electron has no lighter charged particles to decay into, and the neutron undergoes beta decay without violating baryon number conservation, the proton is kinematically allowed to decay into other particles such as a positron plus a neutral pion or a neutrino plus a positively charged pion. But the latter has never been observed. 
On the other hand, baryon number violation is one of the three Sakharov conditions~\cite{Sakharov:1967dj} to explain the overwhelming dominance of matter over antimatter in the current Universe. It is, thus, of fundamental importance to examine the fate of the baryon number, and baryon number violating (BNV) nucleon decays are the most feasible processes that can be searched for in terrestrial laboratories. 

There is a long history of experimental endeavors that date back to the latter half of the 20th century. Large fiducial mass experiments such as 
IMB~\cite{Irvine-Michigan-Brookhaven:1983iap}; 
SNO+~\cite{SNO:2018ydj}; 
KamLAND~\cite{KamLAND:2015pvi};
Kamiokande~\cite{Hirata:1988ad}; and its upgrade, Super-Kamiokande~\cite{Takhistov:2016eqm},
have investigated various characteristic nucleon decays, yielding null results that have imposed stringent limits on their occurrence. The next generation of neutrino experiments~\cite{Dev:2022jbf}, including DUNE~\cite{DUNE:2016evb,DUNE:2020ypp}, Hyper-Kamiokande~\cite{Hyper-Kamiokande:2018ofw}, JUNO~\cite{JUNO:2015zny}, and THEIA~\cite{Theia:2019non}, are expected to further enhance sensitivity to nucleon decays. 
It is important that all possible decay channels should be attempted, exotic as well as conventional, and this requires a complete analysis of relevant interactions at both the effective and fundamental levels. 

BNV nucleon decays are rare, if they exist at all. This implies that baryon number violations must originate from physics at a very high energy scale. Indeed, baryon number violation is a common prediction in theories beyond the standard model (SM), such as grand unified theories~\cite{Georgi:1974sy,Sakai:1981pk,Hisano:1992jj}. Effective field theory (EFT) provides a systematic and consistent framework to relate the physical origin of BNV 
at high energy to the experimentally sought processes at low energy. Most importantly, with minimal assumptions, it does not rely on the details of high energy physics and is, thus, universal in a broad sense. 

In this Letter, we investigate general BNV interactions of the octet baryons that violate the baryon number by one unit ($|\Delta B|=1$), and thus ,contribute to BNV nucleon decays. We explore BNV interactions in the low energy effective field theory (LEFT) that involve three light quarks without being acted upon by a derivative. By matching them to those in chiral perturbation theory (ChPT), we establish BNV effective interactions for the octet baryons and pseudoscalars, neutrinos and charged leptons, and the photon, plus any new light particles beyond the SM, and thus, put the study of all possible BNV nucleon decays on the same footing.

We find that the above triple-quark structures are classified into four irreducible representations (irreps), plus their chirality partners, under the chiral group $G_\chi=\rm SU(3)_\tL\otimes SU(3)_\tR$ of QCD: 
$\pmb{8}_{\tt L(R)}\otimes \pmb{1}_{\tt R(L)}$, 
$\bar{\pmb{3}}_{\tt L(R)}\otimes \pmb{3}_{\tt R(L)}$, 
$\pmb{6}_{\tt L(R)}\otimes \pmb{3}_{\tt R(L)}$, 
$\pmb{10}_{\tt L(R)}\otimes \pmb{1}_{\tt R(L)}$, where the subscripts ${\tt L},{\tt R}$ refer to the left- and right-handed chirality of quarks. 
While the first half of the irreps were known in the 1980s~\cite{Wilczek:1979hc,Ellis:1979hy,Weinberg:1979sa,Weinberg:1980bf,Abbott:1980zj,Kaymakcalan:1983uc} in the form of dimension-6 (dim-6) operators, the other half are identified for the first time. A ChPT was built in 
Ref.\,\cite{Claudson:1981gh} based on the irreps 
$\pmb{8}_{\tt L(R)}\otimes \pmb{1}_{\tt R(L)}$ and 
$\bar{\pmb{3}}_{\tt L(R)}\otimes \pmb{3}_{\tt R(L)}$.  
Since then, nearly all phenomenological analyses of BNV nucleon decays have been based on these results. 
In particular, these chiral interactions have been extensively employed in recent studies of various nucleon decay modes from the EFT perspective~\cite{Hambye:2017qix,Fonseca:2018ehk, Helo:2019yqp,Heeck:2019kgr,Girmohanta:2019xya,He:2021sbl,Dorsner:2022twk,Beneito:2023xbk,Gargalionis:2024nij,Fajfer:2023gfi,Li:2024liy,Li:2025slp,Oosterhof:2021lvt,Oosterhof:2022ljc}. 
As we show in this Letter, the new structures, $\pmb{6}_{\tt L(R)}\otimes \pmb{3}_{\tt R(L)}$ and $\pmb{10}_{\tt L(R)}\otimes \pmb{1}_{\tt R(L)}$, 
naturally appear in higher-dimension operators. We find that they have nontrivial chiral realizations in ChPT, 
operate ($\pmb{6}_{\tt L(R)}\otimes \pmb{3}_{\tt R(L)}$) at the same chiral order as those previously identified ($\pmb{8}_{\tt L(R)}\otimes \pmb{1}_{\tt R(L)}$ and 
$\bar{\pmb{3}}_{\tt L(R)}\otimes \pmb{3}_{\tt R(L)}$)~\cite{Claudson:1981gh}, 
and thus, may significantly influence nucleon decays. 
They are especially important for exotic nucleon decays that are either prohibited at the dim-6 order or involve new light particles.

\vspace{0.1cm}{\bf General BNV triple-quark interactions.}
In the LEFT framework or its extensions with new light fields, we find that a general BNV operator involving three quarks ($q=u,d,s$) without being acted upon by a covariant derivative can always be expressed in terms of the following four structures, 
\begin{subequations}
\label{eq:3qoperator}
\begin{align}
{\cal O}_{a}^{yzw} & = (\overline{\Psi_{a}} q_{\tL, y}^\alpha) (\overline{ q_{\tL, z}^{\beta \C} }  q_{\tL, w}^\gamma )\epsilon_{\alpha \beta \gamma},
\\
{\cal O}_{b}^{yzw} & =(\overline{\Psi_{b}} q_{\tR, y}^\alpha) (\overline{ q_{\tL, z}^{\beta \C} }  q_{\tL, w}^\gamma ) \epsilon_{\alpha \beta \gamma},
\\
{\cal O}_{c}^{yzw} & = (\overline{\Psi_{c,\mu}} q_{\tL,\{y}^\alpha) (\overline{ q_{\tL, z\} }^{\beta \C} } \gamma^\mu q_{\tR, w}^\gamma ) \epsilon_{\alpha \beta \gamma},
\\
{\cal O}_{d}^{yzw} & = (\overline{\Psi_{d,\mu\nu}} q_{\tL,\{y}^\alpha) (\overline{ q_{\tL, z }^{\beta \C} } \sigma^{\mu\nu} q_{\tL, w\}}^\gamma ) \epsilon_{\alpha \beta \gamma},
\end{align} 
\end{subequations}
and their chirality partners with $\tL \leftrightarrow \tR$. 
Here $y,z,w=1,2,3$ are light quark flavor indices, with $q_{1,2,3}=u,d,s$. The curly brackets indicate symmetrization in all flavor indices of like-chirality fields included, viz.,
$A_{\{y}B_{z\}} \equiv (1/2)(A_{y} B_{z} + A_{z}B_{y})$
and $A_{\{y}B_z C_{w\}} \equiv (1/6) [ A_{y}B_{z} C_w + \mbox{5 perms of}(y,z,w)]$. 
The conjugate fermion-type external fields $\overline{\Psi_{a}},\overline{\Psi_{b}},\overline{\Psi_{c,\mu}},\overline{\Psi_{d,\mu\nu}}$ are combination of non-QCD fields that can be identified for each specific operator.
The completeness and independence of the structures in \cref{eq:3qoperator} can be proved using non-standard Fierz identities provided in~\cite{Liao:2016hru,Liao:2020jmn}
(see Supplemental Material for examples). 

We now analyze the transformation properties of the structures in \cref{eq:3qoperator} and their chirality partners under the chiral symmetry group $G_\chi$. 
The triple-quark sector in each structure corresponds to the following irrep respectively, 
\begin{subequations}
\label{eq:Nyzw}
\begin{align}
{\cal N}_{yzw}^{\tL\tL} & =  q_{\tL, y}^\alpha (\overline{ q_{\tL, z}^{\beta \C} } q_{\tL, w}^\gamma )\epsilon_{\alpha \beta \gamma}\in \pmb{8}_\tL \otimes  \pmb{1}_\tR , 
\\
{\cal N}_{yzw}^{\tR\tL} & = q_{\tR, y}^\alpha (\overline{ q_{\tL, z}^{\beta \C} } q_{\tL,w}^\gamma)\epsilon_{\alpha \beta \gamma} \in 
\bar{\pmb{3}}_\tL \otimes \pmb{3}_\tR , 
\\
{\cal N}_{yzw}^{\tL\tR,\mu} & = q_{\tL,\{y}^\alpha (\overline{ q_{\tL, z\}}^{\beta \C} } \gamma^\mu q_{\tR,w}^\gamma)\epsilon_{\alpha \beta \gamma}
\in \pmb{6}_\tL \otimes \pmb{3}_\tR,
\\
{\cal N}_{yzw}^{\tL\tL,\mu\nu} & =  q_{\tL, \{y}^\alpha (\overline{ q_{\tL, z}^{\beta \C} } \sigma^{\mu\nu} q_{\tL, w\} }^\gamma )\epsilon_{\alpha \beta \gamma}\in \pmb{10}_\tL \otimes  \pmb{1}_\tR, 
\end{align}
\end{subequations} 
plus those with $\tL\leftrightarrow\tR$.  
Note that  ${\cal N}_{yzw}^{\tL\tL}$ has no singlet component, i.e., $\epsilon_{yzw} {\cal N}_{yzw}^{\tL\tL}=0$, due to the Fierz identity.
Thus, we may treat ${\cal N}_{uds}^{\tL\tL}$ as a redundant operator in favor of ${\cal N}_{dsu}^{\tL\tL}$ and ${\cal N}_{sud}^{\tL\tL}$. 

When restricted to dim-6 operators in LEFT, $\overline{\Psi_{a,b}}=\bar\psi$, $\overline{\Psi_{c,\mu}}=\bar\psi\gamma_\mu$, and $\overline{\Psi_{d,\mu\nu}}=\bar\psi\sigma_{\mu\nu}$,
where $\psi$ is a charged lepton or neutrino field. The operators ${\cal O}_{a,b}^{yzw}$ corresponding to the structures ${\cal N}_{yzw}^{\tL\tL,\tR\tL}$ are the well-known dim-6 operators in the LEFT~\cite{Jenkins:2017dyc}, while the operators ${\cal O}_{c,d}^{yzw}$ associated with the structures ${\cal N}_{yzw}^{\tL\tR,\mu},~{\cal N}_{yzw}^{\tL\tL,\mu\nu}$ can be shown to vanish by Fierz identities and flavor symmetries. Thus, the new structures ${\cal N}_{yzw}^{\tL\tR,\mu}$ and ${\cal N}_{yzw}^{\tL\tL,\mu\nu}$ first appear as dim-7 and higher operators~\cite{Liao:2020zyx,Murphy:2020cly,Li:2020tsi}. 
For example, the BNV dim-7 operators ${\cal O}_{\bar LQddD}$ in the standard model EFT (SMEFT)~\cite{Liao:2016hru} and ${\cal O}_{QdQND}$ in the SMEFT extended with light sterile neutrinos ($\nu$SMEFT)~\cite{Liao:2016qyd} are essentially 
related to the irreps $\pmb{6}_{\tL(\tR)}\otimes \pmb{3}_{\tR(\tL)}$. 
By using the Fierz identities and integration by parts, these operators can indeed be transformed into the structure given by ${\cal O}_{c}^{yzw}$. 
For the irreps $\pmb{10}_{\tL(\tR)}\otimes \pmb{1}_{\tR(\tL)}$, 
there is no dim-7 LEFT operator, since the only possibility for the non-QCD part of ${\cal O}_{d}^{yzw}$ is $\overline{\Psi_{d,\mu\nu}}=\overline{D_\mu \psi}\gamma_\nu$, which, however, yields a vanishing ${\cal O}_{d}^{yzw}$ upon applying Fierz identities. Nontrivial contributions start to appear at dim 8. An example is 
$B_{\mu\nu}(\overline{e^{\C}} u_{\alpha} ) (\overline{u^{\C}_{\beta}}\sigma^{\mu\nu} d_{\gamma})\epsilon^{\alpha\beta\sigma}$, 
where $B_{\mu\nu}$ is the hypercharge gauge field strength tensor and $e,u,d$ are the right-handed lepton and quark singlets.  
Currently, there are no phenomenological studies that explore these new structures, although they will result in exotic nucleon decays.
In the isospin space, while the known structures ${\cal N}_{yzw}^{\tL\tL}$ and ${\cal N}_{yzw}^{\tR\tL}$ have isospin components $I=0,~1/2,~1$, the new ones, ${\cal N}_{yzw}^{\tL\tR,\mu}$ and ${\cal N}_{yzw}^{\tL\tL,\mu\nu}$, contain the additional component $I=3/2$. Consequently, nucleon and other BNV decays that change isospin by $3/2$ units can be induced by operators associated with these new structures, e.g., the two-body decays $n\to e^- \pi^+$ and $n\to \mu^- \pi^+$.

To facilitate chiral matching, we follow Ref.\,\cite{Fan:2024gzc} to organize the flavor components of the two known representations in matrix form 
\begin{subequations}
\label{eq:3qpart}
\begin{align}
{\cal N}_{{\bf 8}_\tL\otimes {\bf 1}_\tR}
& =
\begin{pmatrix}
{\cal N}^{\tL\tL}_{uds}  &  {\cal N}^{\tL\tL}_{usu}  & {\cal N}^{\tL\tL}_{uud}  
\\[1pt]
{\cal N}^{\tL\tL}_{dds}  & {\cal N}^{\tL\tL}_{dsu} & {\cal N}^{\tL\tL}_{dud}  
\\[1pt]
{\cal N}^{\tL\tL}_{sds} & {\cal N}^{\tL\tL}_{ssu} & {\cal N}^{\tL\tL}_{sud}
\end{pmatrix},
\\
{\cal N}_{\bar{\pmb{3}}_\tL \otimes \pmb{3}_\tR } & = 
 \begin{pmatrix}
{\cal N}_{uds}^{\tR\tL}  
& {\cal N}_{usu}^{\tR\tL} 
& {\cal N}_{uud}^{\tR\tL} 
\\[1pt]
{\cal N}_{dds}^{\tR\tL}  
& {\cal N}_{dsu}^{\tR\tL} 
& {\cal N}_{dud}^{\tR\tL} 
\\[1pt] 
{\cal N}_{sds}^{\tR\tL}  
& {\cal N}_{ssu}^{\tR\tL} 
& {\cal N}_{sud}^{\tR\tL}
 \end{pmatrix}.
\end{align}
\end{subequations}
Together with ${\cal N}_{yzw}^{\tL\tR,\mu}$ and ${\cal N}_{yzw}^{\tL\tL,\mu\nu}$ in \cref{eq:Nyzw}, 
they transform under $(\hat L,\hat R) \in G_\chi$ as 
\begin{subequations}
\begin{align}
{\cal N}_{\pmb{8}_\tL \otimes \pmb{1}_\tR} &\rightarrow
\hat L {\cal N}_{\pmb{8}_\tL \otimes \pmb{1}_\tR} \hat L^\dagger, 
\\%
{\cal N}_{\bar{\pmb{3}}_\tL \otimes \pmb{3}_\tR } &\rightarrow
\hat R {\cal N}_{  \bar{\pmb{3}}_\tL  \otimes \pmb{3}_\tR} \hat L^\dagger,
\\%
{\cal N}_{yzw}^{\tL\tR,\mu} 
&\rightarrow
\hat L_{yy'} \hat L_{zz'} \hat R_{w w'}  
{\cal N}_{y'z'w'}^{\tL\tR,\mu},
\\%
{\cal N}_{yzw}^{\tL\tL,\mu\nu} &\rightarrow
\hat L_{y y'} \hat L_{z z'} \hat L_{w w'}
{\cal N}_{y'z'w'}^{\tL\tL,\mu\nu} . 
\end{align} 
\end{subequations} 
The transformation rules for chirality partners are obtained from the above by swapping $\tL (\hat L)$ with $\tR (\hat R)$. 

Before embarking on chiral matching of the LEFT operators, we find it important to take into account their Lorentz properties.
While the known structures correspond to the simple chiral spinor representations
${\cal N}_{\pmb{8}_\tL \otimes \pmb{1}_\tR},~{\cal N}_{\pmb{3}_\tL 
\otimes \bar{\pmb{3}}_\tR } \in (1/2,0)$ 
and
${\cal N}_{\pmb{1}_\tL \otimes \pmb{8}_\tR},~{\cal N}_{\bar{\pmb{3}}_\tL 
\otimes \pmb{3}_\tR} \in (0,1/2)$ 
under the Lorentz algebra $\mathfrak{su}_l(2)\oplus\mathfrak{su}_r(2)$, the new structures exhibit characteristics of a spin-3/2 field: 
\begin{subequations}
\begin{align}
& {\cal N}_{yzw}^{\tL\tR,\mu} 
\in \big(1,1/2\big), &
& {\cal N}_{yzw}^{\tR\tL,\mu} 
\in \big( 1/2,1\big),
\\
& {\cal N}_{yzw}^{\tL\tL,\mu\nu} 
\in \big( 3/2, 0\big), &
& {\cal N}_{yzw}^{\tR\tR,\mu\nu} 
\in \big(0, 3/2\big),
\end{align}
\end{subequations}
which complicates their chiral matching.
The vanishing of dim-6 operators associated with the new structures is equivalent to the following algebraic identities, 
\begin{align}
  \gamma_\mu {\cal N}_{yzw}^{\tL\tR(\tR\tL),\mu}
=0, \quad
\gamma_\mu {\cal N}_{yzw}^{\tL\tL(\tR\tR),\mu\nu}
=0.
\end{align}
We will demonstrate that considering these irreducible Lorentz representations and algebraic identities is essential for achieving consistent chiral matching. 

For each generic operator ${\cal O}_i^{yzw}$ ($i=a,b,c,d$), we denote its corresponding Wilson coefficient (WC) by $C_i^{yzw}$, which retains the same flavor symmetry as the operator. Furthermore, for each interaction term $C_i^{yzw} {\cal O}_i^{yzw}$, we denote the product of the nonquark field $\overline{\Psi}$ and the WC $C_i^{yzw}$ as a spurion field ${\cal P}_{yzw}^i$. In a manner similar to \cref{eq:3qpart}, we introduce the spurion fields in matrix form, ${\cal P}_{\pmb{8}_\tL \otimes \pmb{1}_\tR,\,\pmb{3}_\tL \otimes \bar{\pmb{3}}_\tR}$ for ${\cal N}_{\pmb{8}_\tL \otimes \pmb{1}_\tR,\,\bar{\pmb{3}}_\tL \otimes \bar{\pmb{3}}_\tR }$,
and denote the spurion fields of ${\cal N}_{yzw}^{\tL\tR,\mu}$ and ${\cal N}_{yzw}^{\tL\tL,\mu\nu}$ 
by ${\cal P}_{yzw}^{\tL\tR,\mu}$ and ${\cal P}_{yzw}^{\tL\tL,\mu\nu}$, where the last one is also antisymmetric in the Lorentz indices $\mu,\,\nu$. We have attributed the $(1,1)$ element (${\cal P}^{\tL\tL}_{uds}$) of $ {\cal P}_{\pmb{8}_\tL \otimes \pmb{1}_\tR}$ to its $(2,2)$ and $(3,3)$ elements (${\cal P}^{\tL\tL}_{dsu}$ and ${\cal P}^{\tL\tL}_{sud}$) by treating ${\cal N}^{\tL\tL}_{uds}$ as redundant, and thus, set ${\cal P}^{\tL\tL}_{uds}=0$. 

It proves convenient to employ the spurion technique for chiral matching. In this technique, the spurion field ${\cal P}$ is assumed to transform in a manner such that the corresponding interaction ${\cal PN}$ is invariant under all symmetries of QCD in the chiral limit, including chiral, Lorentz, baryon-number, and discrete symmetries of charge conjugation ($C$), parity ($P$), and time reversal ($T$). We thus assign the following chiral transformation rules to the spurion fields, 
\begin{subequations}
\label{eq:chitran}
\begin{align}
{\cal P}_{\pmb{8}_\tL \otimes \pmb{1}_\tR}
& \to
\hat L {\cal P}_{\pmb{8}_\tL \otimes \pmb{1}_\tR} \hat L^\dagger,
\\%
{\cal P}_{\pmb{3}_\tL \otimes \bar{\pmb{3}}_\tR } 
& \to  
\hat L {\cal P}_{ \pmb{3}_\tL \otimes \bar{\pmb{3}}_\tR} \hat R^\dagger,
\\%
{\cal P}_{yzw}^{\tL\tR,\mu}
& \to
\hat L^*_{yy'} \hat L^*_{zz'} \hat R_{ww'}^*  
{\cal P}_{y'z'w'}^{\tL\tR,\mu},
\\%
{\cal P}_{yzw}^{\tL\tL,\mu\nu}
& \to \hat L_{yy'}^* \hat L_{zz'}^* \hat L_{ww'}^*
{\cal P}_{y'z'w'}^{\tL\tL,\mu\nu},
\end{align}
\end{subequations}
and similar ones for their chirality partners with $\tL \leftrightarrow \tR$. 
The general $\Delta B=1$ LEFT Lagrangian can now be compactly expressed as 
\begin{align}
{\cal L}_{q^3}^{\slashed{B}} & = 
{\rm Tr} \big[  
  {\cal P}_{\pmb{8}_\tL \otimes \pmb{1}_\tR }
  {\cal N}_{\pmb{8}_\tL \otimes \pmb{1}_\tR } 
+ {\cal P}_{ \pmb{1}_\tL \otimes \pmb{8}_\tR }  
  {\cal N}_{  \pmb{1}_\tL \otimes \pmb{8}_\tR }
  \big]  
\nonumber 
\\
& + {\rm Tr} \big[ 
  {\cal P}_{\pmb{3}_\tL \otimes \bar{\pmb{3}}_\tR }
  {\cal N}_{\bar{\pmb{3}}_\tL \otimes \pmb{3}_\tR } 
+ {\cal P}_{\bar{\pmb{3}}_\tL \otimes \pmb{3}_\tR }
  {\cal N}_{\pmb{3}_\tL \otimes \bar{\pmb{3}}_\tR }
 \big] 
\nonumber 
\\
& + \big[
{\cal P}_{yzw}^{\tL\tR,\mu}
{\cal N}_{yzw,\mu}^{\tL\tR}
+ {\cal P}_{yzw}^{\tR\tL,\mu}
{\cal N}_{yzw,\mu}^{\tR\tL}
\big]
\nonumber 
\\%
& + \big[
{\cal P}_{yzw}^{\tL\tL,\mu\nu}
{\cal N}_{yzw,\mu\nu}^{\tL\tL}
+ 
{\cal P}_{yzw}^{\tR\tR,\mu\nu}
{\cal N}_{yzw,\mu\nu}^{\tR\tR}
\big]
 +\hc,
\label{eq:q3LEFT1}
\end{align}
where the trace is over the three-flavor space and the repeated indices $y,z,w$ are summed over.
Any analogous BNV interactions expressed in other bases can always be transformed into the canonical form in \cref{eq:q3LEFT1} 
to facilitate chiral matching. 

The effective operators ${\cal O}_i^{yzw}$ in \cref{eq:3qoperator} are typically matched at the electroweak (EW) scale $\Lambda_{\tt EW}\approx160~\rm GeV$ with those in SMEFT or its extensions. Consequently, their contributions to nucleon decays below the chiral symmetry breaking scale $\Lambda_{\chi}\approx 1.2~\rm GeV$ are influenced by the renormalization group evolution (RGE) effect, with the leading one coming from one-loop QCD corrections. Using the RGE equations in Supplemental Material, we find 
$C_{a,b}^{yzw}(\Lambda_\chi)\approx 1.32\,C_{a,b}^{yzw}(\Lambda_{\tt EW})$,
$C_c^{yzw}(\Lambda_\chi)\approx 0.91\,C_c^{yzw}(\Lambda_{\tt EW})$, and 
$C_d^{yzw}(\Lambda_\chi)\approx 0.76\,C_d^{yzw}(\Lambda_{\tt EW})$.

\vspace{0.1cm}{\bf Chiral matching.}
To obtain the hadronic counterparts of the quark-level interactions in \cref{eq:q3LEFT1}, a consistent approach is to work with ChPT, which can systematically account for interactions of both the octet pseudoscalar mesons and baryons. In ChPT, the meson and baryon octet fields are organized in matrix form, 
\begin{subequations}
\begin{align}
\Sigma(x) & = \xi^2(x) = \exp\Big(\frac{i\sqrt{2}\Pi(x)}{F_0}\Big),  \\
\Pi(x) & =   
\begin{pmatrix}
\frac{\pi^0}{\sqrt{2}}+\frac{\eta}{\sqrt{6}} & \pi^+ & K^+
\\
\pi^- & -\frac{\pi^0}{\sqrt{2}}+\frac{\eta}{\sqrt{6}} & K^0
\\
K^- & \bar{K}^0 & -\sqrt{\frac{2}{3}}\eta
\end{pmatrix},
\\
B(x) &=
\begin{pmatrix}
{\Sigma^{0}\over \sqrt{2}}+{\Lambda^0 \over \sqrt{6}}  & \Sigma^+ & p \\
\Sigma^- & -{\Sigma^{0} \over \sqrt{2}}+{\Lambda^0 \over \sqrt{6}} &  n \\ 
\Xi^- & \Xi^0 & - \sqrt{2\over 3}\Lambda^0
\end{pmatrix},
\end{align}
\end{subequations}
where $F_0=(86.2 \pm 0.5)~\rm MeV$ is the pion decay constant in the chiral limit. Their chiral transformation rules are  
$\Sigma \to \hat L \Sigma \hat R^\dagger$, 
$ B \to \hat h B \hat h^\dagger$, 
$\xi \to \hat L \xi \hat h^\dagger = \hat h \xi \hat R^\dagger$, 
where the matrix $\hat h$ is a function of $\hat L,~\hat R$, and $\xi$. For convenience, we note that 
$\xi B \xi \to 
\hat L(\xi B \xi) \hat R^\dagger$,  
$\xi^\dagger B \xi^\dagger \to 
\hat R (\xi^\dagger B \xi^\dagger) \hat L^\dagger$, 
$\xi B \xi^\dagger \to 
\hat L (\xi B \xi^\dagger) \hat L^\dagger$,
and $\xi^\dagger B \xi \to 
\hat R (\xi^\dagger B \xi) \hat R^\dagger$. 

The chiral building blocks corresponding to the LEFT interactions in \cref{eq:q3LEFT1} consist of the spurion fields, the meson and baryon octet fields, and their derivatives. To identify their leading chiral counterparts, we adopt the chiral derivative power counting scheme in which 
$\{\Sigma, \xi,B,D_\mu B\}\sim {\cal O}(p^0)$ and 
$D_\mu\Sigma \sim {\cal O}(p^1)$. The covariant derivatives are defined as $D_\mu \Sigma = \partial_\mu \Sigma -il_{\mu}\Sigma+i\Sigma r_{\mu}$, 
$D_\mu B = \partial_\mu B + [\Gamma_\mu ,B]$, where the external sources $l_{\mu}$ and $r_{\mu}$ are traceless matrices in the flavor space and $\Gamma_{\mu}=(1/2)\left[\xi(\partial_{\mu}-ir_{\mu})\xi^{\dagger}+\xi^{\dagger}(\partial_{\mu}-il_{\mu})\xi\right]$. These covariant derivatives transform as $B$ and $\Sigma$ themselves. It is important that we must account for Lorentz as well as chiral properties on chiral matching.  
We find that, for the chiral irreps $\pmb{6}_{\tL(\tR)}\otimes \pmb{3}_{\tR(\tL)}$, the $(1,1/2)$ and $(1/2,1)$ components of a general vector-spinor object, such as $\partial^\nu B$, is extracted by the projectors
\begin{align}
{\Gamma}_{\mu\nu}^{\tt L,R} \equiv \Big(g_{\mu\nu} - {1\over 4} \gamma_\mu \gamma_\nu\Big)P_{\tt L,R},    
\end{align}
which satisfy the relations
$ {\Gamma}_{\mu\rho}^{\tt L,R} 
{\Gamma}_{~\nu}^{\tt L,R~\rho} 
={\Gamma}_{\mu\nu}^{\tt L,R} $ and
$\gamma^\mu {\Gamma}_{\mu\nu}^{\tt L,R}=0$.
On the other hand, for the chiral irreps $\pmb{10}_{\tL(\tR)}\otimes \pmb{1}_{\tR(\tL)}$, we have to resort to the following projectors~\cite{DelgadoAcosta:2015ypa},
\begin{align}
\hat\Gamma_{\mu\nu\alpha\beta}^{\tt L,R}  \equiv
{1\over 24} \left( 
2 \{\sigma_{\mu\nu},\sigma_{\alpha\beta}\}
- 
[\sigma_{\mu\nu},\sigma_{\alpha\beta}]
\right) P_{\tt L,R},
\label{eq:LorentzP}
\end{align}
to extract the irreducible $(3/2,0)$ and $(0,3/2)$ components of a tensor-spinor object such as $\partial^\alpha B \partial^\beta \Sigma$. 
These projectors satisfy similar relations, 
$\hat\Gamma_{\mu\nu \rho\sigma}^{\tt L,R} 
\hat\Gamma_{~~\alpha\beta}^{\tt L,R~\rho\sigma}=
\hat\Gamma_{\mu\nu \alpha\beta}^{\tt L,R}$
and $\gamma^\mu \hat\Gamma_{\mu\nu \alpha\beta}^{\tt L,R} =0$.

We can now construct, order by order, the chiral interactions for the LEFT interactions in \cref{eq:q3LEFT1} that satisfy all required symmetries and Lorentz properties. 
We find that each of the chiral irreps 
$\bar{\pmb{3}}_{\tL(\tR)} \otimes \pmb{3}_{\tR(\tL)}$, 
$\pmb{8}_{\tL(\tR)} \otimes \pmb{1}_{\tR(\tL)}$, and
$\pmb{6}_{\tL(\tR)} \otimes \pmb{3}_{\tR(\tL)}$  has a unique chiral realization which appears at the same chiral order. 
They can be succinctly parametrized as follows: 
\begin{align}
{\cal L}_{B}^{\slashed{B},0} &=
c_1 {\rm Tr}\big[ 
{\cal P}_{  \bar{\pmb{3}}_\tL \otimes \pmb{3}_\tR} \xi B_\tL \xi -
{\cal P}_{\pmb{3}_\tL \otimes \bar{\pmb{3}}_\tR} \xi^\dagger B_\tR \xi^\dagger 
 \big]
\nonumber
\\
& + c_2 {\rm Tr}\big[ 
{\cal P}_{\pmb{8}_\tL \otimes \pmb{1}_\tR}\xi B_\tL \xi^\dagger
- {\cal P}_{ \pmb{1}_\tL \otimes  \pmb{8}_\tR} \xi^\dagger B_\tR \xi
\big] 
\nonumber
\\
& + {c_3 \over \Lambda_\chi} \big[ 
{\cal P}_{yzi}^{\tL\tR,\mu}
{\Gamma}_{\mu\nu}^{\tt L} 
(\xi i D^\nu B_\tL \xi)_{yj}
\Sigma_{zk} \epsilon_{ijk}
\nonumber
\\
& - {\cal P}_{yzi}^{\tR\tL,\mu}
{\Gamma}_{\mu\nu}^{\tt R}
(\xi^\dagger i D^\nu B_\tR  \xi^\dagger)_{yj}
\Sigma^*_{kz} \epsilon_{ijk} \big]
+\hc,
\label{eq:chiB0}
\end{align} 
where all of the indices $y,z$ and $i,j,k$ are summed over the three flavors $u,d,s$, and $B_{\tL(\tR)}\equiv P_{\tL(\tR)}B$. The relative minus sign in each term arises from parity consideration. 
The parameters $c_{1-3}$ (and $c_4,\tilde c_3,\tilde c_4$ below) are unknown low energy constants (LECs) that can be estimated based on the naive dimensional analysis (NDA) or calculated by lattice QCD; 
see Supplemental Material for details. Note that the factor $\Lambda_\chi^{-1}$ in the $c_3$ term is balanced by the derivative acting on the baryon field according to chiral power counting.

For the totally symmetric irreps 
$\pmb{10}_{\tL(\tR)}\otimes \pmb{1}_{\tR(\tL)}$, the leading chiral matching is also unique, but it must involve a derivative acting on the octet mesons, and thus, appears only at a higher chiral order than the other irreps. Using the projectors in \cref{eq:LorentzP},  we parametrize it as follows:
\begin{align}
{\cal L}_{B}^{\slashed{B},1} & =  
{c_4 \over \Lambda_\chi^2} \big[ 
{\cal P}_{yzw}^{\tL\tL,\mu\nu}
\hat\Gamma_{\mu\nu \alpha\beta}^{\tt L} 
(\xi D^{\alpha} B_\tL \xi)_{yi}
\Sigma_{zj}(D^{\beta}\Sigma)_{wk} \epsilon_{ijk}
\nonumber
\\
& - {\cal P}_{yzw}^{\tR\tR,\mu\nu} 
\hat\Gamma_{\mu\nu \alpha\beta}^{\tt R}  
(\xi^\dagger D^{\alpha} B_\tR \xi^\dagger)_{yi}
\Sigma^*_{jz} (D^{\beta}\Sigma)^*_{kw} \epsilon_{ijk}\big]
\nonumber
\\
&+\hc.
\label{eq:chiB1}
\end{align}
Since the spin-3/2 decuplet baryons share the same symmetry properties as the triple quarks in $\pmb{10}_{\tL(\tR)}\otimes \pmb{1}_{\tR(\tL)}$, it looks natural to include them. Denoting them by the vector-spinor field $T^\mu_{ijk}$, 
which is totally symmetric in the flavor indices $ijk$ and transforms under $G_\chi$ as $T^\mu_{ijk}\to \hat{h}_{ii'} \hat{h}_{jj'} \hat{h}_{kk'} T^\mu_{i'j'k'}$, the leading BNV terms are
\begin{align}
{\cal L}_{T}^{\slashed{B},0}&=\tilde{c}_3\big[{\cal P}^{\tL\tR,\mu}_{yzw} \Gamma^{\tL}_{\mu\nu} T_{ijk}^{\nu} \xi_{yi} \xi_{zj} \xi^*_{kw} 
\nonumber
\\
& -{\cal P}^{\tR\tL,\mu}_{yzw} \Gamma^{\tR}_{\mu\nu} T_{ijk}^{\nu} \xi^*_{iy} \xi^*_{jz} \xi_{wk}\big]
\nonumber
\\
&+{\tilde{c}_4 \over \Lambda_\chi}\big[{\cal P}^{\tL\tL,\mu\nu}_{yzw} \hat\Gamma^{\tL}_{\mu\nu\alpha\beta} i D^\alpha T_{ijk}^{\beta} \xi_{yi} \xi_{zj} \xi_{wk}
\nonumber
\\
&-{\cal P}^{\tR\tR,\mu\nu}_{yzw} \hat\Gamma^{\tR}_{\mu\nu\alpha\beta} i D^\alpha T_{ijk}^{\beta} \xi^*_{iy} \xi^*_{jz} \xi^*_{kw}\big]+\hc.
\label{eq:chiT0}
\end{align}
These may contribute a decuplet pole term to BNV nucleon decays with the help of a conventional nucleon-decuplet-meson interaction. In Supplemental Material, we show an example of such a contribution.

Equations (\ref{eq:chiB0}) and (\ref{eq:chiB1}) are our main results in this work. The irreps 
$\bar{\pmb{3}}_{\tL(\tR)} \otimes \pmb{3}_{\tR(\tL)}$ and 
$\pmb{8}_{\tL(\tR)} \otimes \pmb{1}_{\tR(\tL)}$
are common in the literature~\cite{Claudson:1981gh}, and they are closely related to the dim-6 BNV operators in the LEFT or SMEFT. However, our construction is more universal. With the aid of general spurion fields, our formalism allows for broader applicability that extends beyond the dim-6 order.

\vspace{0.1cm}{\bf Some phenomenological applications.}
The Lagrangians in \cref{eq:chiB0,eq:chiB1} generate interactions of the type $M^n({\cal P}B)$, involving a single baryon $B$, $n$ pseudoscalar mesons $M$, and a spurion ${\cal P}$, which may contain one or more nonhadronic fields. In what follows, we restrict ourselves to the nucleon ($\textsc n=p,n$) decays. Without a pseudoscalar in the decay product, all irreps except $\pmb{10}_{\tL(\tR)}\otimes \pmb{1}_{\tR(\tL)}$ may contribute.
With a pseudoscalar in the final state, the irreps $\pmb{10}_{\tL(\tR)}\otimes \pmb{1}_{\tR(\tL)}$ start to contribute.
In this case, there are generally two types of contributions: a contact term due to local interactions in \cref{eq:chiB0,eq:chiB1} and a pole term due to the exchange of an octet or decuplet baryon with the help of a usual $BBM$ or $BTM$ vertex.
For the purpose of illustration, we discuss briefly a few BNV nucleon decays that are induced by the contact terms due to the new triple-quark structures, while leaving a more complete phenomenological analysis for future work.

Consider first the decays $n\to\ell^-\pi^+$ with $\ell=e,~\mu$ that change isospin by $3/2$ units. The leading contribution is from the $c_3$ terms in \cref{eq:chiB0} associated with the irreps $\pmb{6}_{\tL(\tR)}\otimes \pmb{3}_{\tR(\tL)}$.
Such interactions can be induced from dim-7 LEFT operators that in turn may descend from the following dim-7 SMEFT operator upon applying Fierz identities and equations of motion: 
\begin{align}
{\cal O}_{DLddQ}^{prst}&= 
(\overline{i D_\mu L_p} d^\alpha_{\{r})
(\overline{d^{\beta \C}_{s\} }}\gamma^\mu Q^\gamma_t )
\epsilon_{\alpha\beta\gamma}. 
\end{align}
This fixes the spurion field ${\cal P}_{ddd}^{\tR\tL,\mu} = - 0.91C_{DLddQ}^{\ell 111}i\partial^\mu \overline{\ell_{\tL}}$. 
We have incorporated the QCD RGE effect and ignored the CKM matrix, 
and the WC at the scale $\Lambda_{\tt EW}$ can be parametrized as $C_{DLddQ}^{\ell 111}= \Lambda_\ell^{-3}$.
The decay rate is found to be
$\Gamma(n\to\ell^-\pi^+)
\sim (5\times 10^9\,\rm GeV/\Lambda_\ell)^6/(10^{31}\,\rm yr)$ for $\ell=e,\mu$, 
where the number in the denominator indicates the current experimental lower bound or sensitivity on the inverse decay width.

As a second example involving only SM particles in the final state, we consider the decays $p\to \ell^+\ell^+\ell'^-$ with $\ell\ell'=e\mu,\mu e$. These are among the most stringently constrained processes in experimental searches for proton decay~\cite{Hambye:2017qix,Super-Kamiokande:2020tor,ParticleDataGroup:2024cfk}. They can be mediated only by dim-9 (and higher) LEFT operators, and thus, are connected again to the irreps $\pmb{6}_{\tL(\tR)}\otimes \pmb{3}_{\tR(\tL)}$. 
To appreciate the importance of these new representations, let us consider the contribution from the following dim-9 LEFT operator, 
\begin{align}
 {\cal O}_{\ell \ell'} & = (\overline{\ell_\tL^\C}\gamma_\mu \ell_\tR)
 (\overline{\ell'_\tR} u_\tL^\alpha)
(\overline{u_\tL^{\beta\C}}\gamma^\mu d_\tR^\gamma) \epsilon_{\alpha\beta\gamma},
\end{align}
with a WC $\Lambda_{\ell\ell'}^{-5}$. 
This yields a spurion field 
${\cal P}_{uu d}^{\tL\tR,\mu} = \Lambda_{\ell\ell'}^{-5} (\overline{\ell_\tL^\C}\gamma^\mu \ell_\tR)
\overline{\ell'_\tR}$ in the $c_3$ term of \cref{eq:chiB0}. 
The decay widths are found to be 
$\Gamma(p\to \ell^+\ell^+\ell'^-)
\sim (3.4\times 10^5{\rm GeV}/\Lambda_{\ell\ell'} )^{10}
/(10^{34}\,\rm yr)$ on neglecting the lepton mass.

When there is a new light particle, it may appear in the decay product of nucleons. We consider two popular examples that may act as dark matter: the axion (or axionlike particle) and the dark photon. Because of a shift symmetry, the axion appears in a derivative form, so that a natural BNV dim-8 LEFT operator is 
\begin{align}
{\cal O}_a = (\partial_\mu a) 
(\overline{e_\tL^\C} u_\tL^\alpha)(\overline{u_\tL^{\beta\C}}\gamma^\mu d_\tR^\gamma)\epsilon_{\alpha\beta\gamma},    
\end{align}
which is, again, in the irrep $\pmb{6}_{\tL}\otimes \pmb{3}_{\tR}$. Denoting its WC as $\Lambda_{a}^{-4}$, this corresponds to the spurion field ${\cal P}_{uu d}^{\tL\tR,\mu} =\Lambda_{a}^{-4}(\partial^\mu a) \overline{e_\tL^\C} $ in the Lagrangian at both quark [\cref{eq:q3LEFT1}] and hadron [\cref{eq:chiB0}] levels. The latter then induces a contribution to the decay $p\to e^+a$.
We find  $\Gamma(p\to e^+ a)
\sim {(1.4 \times 10^7{\rm GeV}/\Lambda_{a}} )^8/(10^{33}\,{\rm yr})$.
The new structures $\pmb{6}_{\tL(\tR)}\otimes \pmb{3}_{\tR(\tL)}$ were overlooked or incorrectly discarded as terms of a higher chiral order than the known ones $\pmb{8}_{\tL(\tR)} \otimes \pmb{1}_{\tR(\tL)}$ and $\bar{\pmb{3}}_{\tL(\tR)} \otimes \pmb{3}_{\tR(\tL)}$ 
in previous works~\cite{Fridell:2023tpb, Li:2024liy}. 

As a final example of the new structures, we consider the field strength tensor of a dark photon coupled to triple quarks and a charged lepton via the dim-8 operator, 
\begin{align}
{\cal O}_{X\ell} = X_{\mu\nu}
(\overline{\ell_\tR} d_\tL^\alpha ) (\overline{d^{\beta\C}_\tL}\sigma^{\mu\nu} d_\tL^{\gamma})\epsilon_{\alpha\beta\gamma},
\end{align}
with a WC $\Lambda^{-4}_{X\ell}$. This interaction belongs to the irrep $\pmb{10}_{\tL}\otimes \pmb{1}_{\tR}$, and it brings in a spurion field ${\cal P}_{ddd}^{\tL\tL,\mu\nu} = \Lambda_{X\ell}^{-4} X^{\mu\nu}\overline{\ell_\tR}$. 
It thus induces the decay $n\to \ell^-\pi^+ X$ ($\ell=e,\,\mu$) 
which changes isospin by 3/2 units and would not be possible by other structures at the same dim-8 order. The decay rate is found to be 
$\Gamma(n\to \ell^-\pi^+ X)
\sim (3\times 10^6\,\rm GeV/\Lambda_{X\ell})^8/(10^{30}\,\rm yr)$.

Many other BNV nucleon decays are possible, for instance, $n\to\nu_{e,\tau}(\bar \nu_{\mu,\tau}) e^-\mu^+,\nu_{\mu,\tau}(\bar \nu_{e,\tau}) \mu^- e^+, 3\nu(\bar\nu)$. For systematic investigations of BNV nucleon decays involving new light particles such as sterile neutrinos or scalar or vector particles, it is advantageous to work in an EFT framework like $\nu$SMEFT~\cite{Liao:2016qyd} and dark sector EFT (DSEFT)~\cite{Liang:2023yta}. 
As a final remark, we illustrate by a model how new structures may be generated at a high energy scale. We introduce three leptoquarks which carry the quantum numbers of the SM gauge group $\rm SU(3)_c\otimes SU(2)_L\otimes U(1)_Y$, 
$\Phi_1(\pmb{3},\pmb{1},-1/3)$, 
$\Phi_2({\pmb 3}, \pmb{2},1/6)$, and 
$\Phi_3(\pmb{3}, \pmb{3},-1/3)$. We impose a $\mathbb{Z}_2$ symmetry under which the electron-type leptons $e_e,~L_e$ and the leptoquarks $\Phi_{1,3}$ are odd while all other fields are even. Due to the flavorful $\mathbb{Z}_2$ symmetry, the model starts to generate BNV SMEFT interactions only at dimension 10 
via the diagram in \cref{fig:BNVdim10}.
Upon spontaneous breaking of the electroweak symmetry, these interactions will lead to dim-9 LEFT operators in the irreps $\pmb{3}_{\tL}\otimes \pmb{6}_{\tR}$ and 
$\pmb{3}_{\tL}\otimes \bar{\pmb{3}}_{\tR}$. 

\begin{figure}[t]
\centering
\begin{tikzpicture}[decoration={markings, 
mark= at position 0.6 with {\arrow{stealth}},
mark= at position 3cm with {\arrow{stealth}}},scale=0.7]
\begin{scope}[shift={(1,1)}]
\draw[postaction={decorate},thick] (0, 0.7) --(-1.0,1.5) node[left]{$u$};
\draw[postaction={decorate},thick, purple] (0, 0.7) -- (1.0,1.5) node[right]{$e_e$};
\draw[dashed,postaction={decorate},ultra thick, purple] (0,-0.3) -- (0,0.7) node[midway,xshift = -10 pt]{$\Phi_{1}$};
\draw[dashed,postaction={decorate}, ultra thick,purple] (0,-0.3) -- (2,-0.3) node[midway, xshift=3pt, yshift = - 8 pt, purple]{$\Phi_3 $};
\draw[postaction={decorate},thick] (2,-0.3)--(3,0.7) node[right]{$Q$};
\draw[postaction={decorate},thick, purple] (2,-0.3)--(3,-1.3) node[right]{$L_e$};
\draw[dashed,postaction={decorate}, ultra thick] (0,-0.3) -- (0,-1.3) node[midway,xshift = -10 pt]{$\Phi_{2}$};
\draw[dashed,postaction={decorate},thick] (0,-0.3) -- (-2,-0.3)node[left]{$H$};
\draw[postaction={decorate},thick] (0,-1.3)--(-1.0,-2.1) node[left]{$d$};
\draw[postaction={decorate},thick](1.0,-2.1)node[right]{$L_{\mu,\tau}$}--(0,-1.3);
\draw[draw=cyan,fill=cyan] (0,-0.3) circle (0.08cm);
\end{scope}
\end{tikzpicture}
\vspace{-1em}
\caption{Diagram that induces dim-10 BNV interactions. $H,~Q,~L$ represent the SM Higgs, left-handed quark and lepton doublets, while $e,u,d$ are the right-handed lepton and quark singlets, respectively.}
\label{fig:BNVdim10}
\end{figure}
\vspace{0.1cm}{\bf Conclusion.}
The study of baryon number violating nucleon decays represents an exhilarating frontier in the exploration of new physics. Many experiments have been designed to search for nucleon decays, highlighting their significance in uncovering fundamental aspects of particle interactions. The general classification of triple-quark interactions and their chiral matching developed in this Letter would find many applications in the search of BNV processes. Notably, we have identified new chiral structures that go beyond the BNV interactions originating from the conventional dim-6 LEFT operators. This opens up new avenues for experimental investigations, including novel nucleon decays involving new light particles that may connect BNV interactions with the nature of dark matter as well as the neutrino mass puzzle. We emphasize that these structures are also pertinent to hyperon and tau lepton BNV decays. 
 
\section*{Acknowledgements}
We thank one anonymous referee for asking us to comment on contributions of the decuplet baryons and another anonymous referee to remind us of Refs.~\cite{Oosterhof:2021lvt,Oosterhof:2022ljc}.
This work was supported 
by Grants 
No.\,NSFC-12035008, 
No.\,NSFC-12247151, 
and No.\,NSFC-12305110. 

\bibliography{references_paper}{}
\bibliographystyle{utphys}

\onecolumngrid
\setcounter{figure}{0}
\renewcommand{\thefigure}{S\arabic{figure}}
\setcounter{equation}{0}
\renewcommand{\theequation}{S\arabic{equation}}

\begin{center}
\vspace{0.1cm}{\bf Supplemental Material}
\end{center}

{\bf{Examples demonstrating completeness and independence of structures in Eq.\,(1)}---}%
For instance, by using the Fierz identity, 
$  (\overline{\psi_{1\tR}}\psi_{2\tL}) (\overline{\psi_{3\tL}^\C}\psi_{4\tL})
= -(\overline{\psi_{1\tR}}\psi_{3\tL}) (\overline{\psi_{2\tL}^\C}\psi_{4\tL})
  -(\overline{\psi_{1\tR}}\psi_{4\tL}) (\overline{\psi_{3\tL}^\C}\psi_{2\tL})$ 
and its variation, operators with antisymmetric flavor indices in two like-chirality quarks in two bilinears, or with a tensor quark bilinear, are not independent but can be transformed into those in \cref{eq:3qoperator},
\begin{align}
&(\overline{\Psi_{c,\mu}} q_{\tL,[y}^\alpha) 
(\overline{ q_{\tL, z]}^{\beta \C} } \gamma^\mu q_{\tR,w}^\gamma)
\epsilon_{\alpha \beta\gamma} 
\nonumber
={1 \over 2} (
\colorbox{gray!20}{$\overline{\Psi_{c,\mu}}\gamma^\mu $} 
q_{\tR,w}^\alpha) 
(\overline{ q_{\tL, z}^{\beta \C} } q_{\tL,y}^\gamma)
\epsilon_{\alpha\beta\gamma} 
\sim{\cal O}_b^{yzw},
\\%
&(\overline{\Psi_{d,\mu\nu}} q_{\tR, y}^\alpha)  
(\overline{ q_{\tL, z}^{\beta \C} } \sigma^{\mu\nu} q_{\tL,w}^\gamma)
\epsilon_{\alpha \beta \gamma} 
= -2 i (
\colorbox{gray!20}{$ \overline{\Psi_{d,\mu\nu}} \gamma^\mu $} 
q_{\tL, \{z}^\alpha)  
(\overline{ q_{\tL, w\}}^{\beta \C}}\gamma^{\nu}q_{\tR, y}^\gamma) 
\epsilon_{\alpha \beta \gamma}
\sim {\cal O}_c^{yzw},
\\%
& (\overline{\Psi_{d,\mu\nu}} q_{\tL, [y}^\alpha)  
(\overline{ q_{\tL, z]}^{\beta \C} } \sigma^{\mu\nu} q_{\tL, w}^\gamma)
\epsilon_{\alpha \beta \gamma} 
\nonumber
= - {1 \over 2} (
\colorbox{gray!20}{$ \overline{\Psi_{d,\mu\nu}}\sigma^{\mu\nu}$} 
q_{\tL, w}^\alpha) 
(\overline{ q_{\tL, y}^{\beta \C}} q_{\tL, z}^\gamma)
\epsilon_{\alpha \beta \gamma}
\sim{\cal O}_a^{yzw},
\end{align}
where 
$A_{[y}B_{z]} \equiv (1/2) (A_{y}B_{z} - A_{z}B_{y})$, and each portion highlighted in gray can be regarded as a single object.

{\bf{One-loop QCD RGE for operators in Eq.\,(1)}---}%
Since these operators belong to distinct chiral irreps, they undergo separate multiplicative renormalization without mixing. Following the method in~\cite{Liao:2019gex}, we obtain the following RGE equations, 
\begin{subequations}
\begin{align}
{dC_{a,b}^{yzw}  \over d\ln \mu} &= - 2 {\alpha_s \over 2\pi} C_{a,b}^{yzw},
\label{eq:8133running}
\\
{dC_c^{yzw}  \over d\ln \mu} &= + {2\over 3}{\alpha_s \over 2\pi} C_c^{yzw},
\\
{dC_d^{yzw}\over d\ln \mu}  &= +2 {\alpha_s \over 2\pi} C_d^{yzw},
\end{align}
\end{subequations}
where $\alpha_s=g_s^2/(4\pi)$ with $g_s$ being the QCD coupling constant. The results in \cref{eq:8133running} for the two usual representations, 
$\pmb{8}_\tL\otimes \pmb{1}_\tR$ and $\bar{\pmb{3}}_\tL\otimes \pmb{3}_\tR$, are consistent with those of the corresponding dim-6 operators given in~\cite{Jenkins:2017dyc}. The same results apply to the WCs of the chirality partner operators. 

{\bf{Naive dimensional analysis for LECs}---}%
The LECs $c_{1,2}$ are identified with $\alpha$ and $\beta$ used in the literature. Recent LQCD calculation leads to $c_1=\alpha=-0.01257(111)\,{\rm GeV}^3$ and $c_2=\beta=0.01269(107)\,{\rm GeV}^3$~\cite{Yoo:2021gql}, respectively. The two terms associated with 
$\pmb{6}_{\tL(\tR)} \otimes \pmb{3}_{\tR(\tL)}$
and $\pmb{10}_{\tL(\tR)} \otimes \pmb{1}_{\tR(\tL)}$
are entirely new. To obtain a first estimation of the LECs $c_{3,4}$, we appeal to the NDA method~\cite{Weinberg:1989dx,Manohar:1983md,Gavela:2016bzc}.
Following Weinberg's approach~\cite{Weinberg:1989dx}, we introduce reduced couplings for both the quark- and hadron-level operators, and then match them to determine the LECs. For an interaction term involving a coupling constant $g$ and a dim-$D$ operator that contains a minimum number $m$ of physical fields, its reduced coupling is defined as $g(4\pi)^{2-m} \Lambda_\chi^{D-4}$. 
For the pure triple-quark sector in \cref{eq:q3LEFT1}, we have $g=1$, $m=3$, and $D=9/2$, resulting in a reduced coupling given by $C_q = (4\pi)^{-1} \Lambda_\chi^{1/2}$. 
For the corresponding hadron-level interactions in \cref{eq:chiB0,eq:chiB1}, we expand the meson matrix to the first nonvanishing order and obtain 
$c_{1,2} B$ ($g=c_{1,2}$, $m=1$, and $D=3/2$),
$(c_{3}/\Lambda_\chi) \partial B$ ($g=c_3/\Lambda_\chi$, $m=1$, and $D=5/2$),
and $\sqrt{2}c_4/(\Lambda_\chi^2 F_0)\partial B \partial M$ ($g=\sqrt{2}c_4/(\Lambda_\chi^2F_0)$, $m=2$, and $D=9/2$), 
where the spurion fields are omitted since they do not participate in chiral matching. Consequently, we find the reduced couplings for the hadron-level operators to be 
$C_{1,2,3} =c_{1,2,3} (4\pi) \Lambda_\chi^{-5/2}$
and $C_4 = (\sqrt{2} c_4/F_0) \Lambda_\chi^{-3/2}$,
respectively. 
By identifying $C_{1-4}=C_q$, we ultimately obtain  
$c_{1,2,3} \sim \Lambda_\chi^3/(4\pi)^2 \approx 0.011\,{\rm GeV}^3$
and 
$c_4 \sim \Lambda_\chi^2 F_0 /(4\pi\sqrt{2}) \approx 0.007\,{\rm GeV}^3$. 
As observed, the NDA estimation yields a value for $c_{1,2}$ that is consistent with the LQCD determination to a high degree of precision. Nevertheless, given that LQCD provides a more accurate estimation of LECs, we strongly encourage the LQCD community to compute these two new LECs $c_{3,4}$. 
Similarly, we estimate by NDA that $\tilde{c}_{3,4} \sim \Lambda_\chi^3/(4\pi)^2 \approx 0.011\,{\rm GeV}^3$.

\begin{figure}[h]
\centering
\begin{tikzpicture}[mystyle,scale=1]
\begin{scope}[shift={(1,1)}]
\draw[f] (0, 0)node[left]{$n$} -- (1.5,0);
\draw[f] (1.5, 0) -- (3,0) node[right]{$\ell^-$};
\draw[v, purple] (1.5,0) -- (2.5,1.2) node[right,yshift = 2 pt]{$X$};
\draw[snar, black] (1.5,0) -- (0.5,1.2) node[right,yshift = 2 pt]{$\pi^+$};
\filldraw [black] (1.38,-0.12) rectangle (1.62,0.12);
\node at (1.5,-1.5) {(a)};
\end{scope}
\end{tikzpicture}
\quad\quad
\begin{tikzpicture}[mystyle,scale=1]
\begin{scope}[shift={(1,1)}]
\draw[f] (0, 0)node[left]{$n$} -- (1.5,0);
\draw[f] (1.5, 0) -- (3,0) node[midway,yshift = - 7pt]{$\Delta^-$};
\draw[f] (3.0, 0) -- (4.5,0) node[right]{$\ell^-$};
\draw[v, purple] (2.9,0) -- (3.9,1.2) node[right,yshift = 2 pt]{$X$};
\draw[snar, black] (1.5,0) -- (2.5,1.2) node[right,yshift = 2 pt]{$\pi^+$};
\filldraw [black] (1.5,0) circle (3pt);
\filldraw [black] (2.88,-0.12) rectangle (3.12,0.12);
\node at (1.5,-1.5) {(b)};
\end{scope}
\end{tikzpicture}
\caption{Feynman diagrams contributing to the decay $n\to \ell^- \pi^+ X$ due to the operator ${\cal O}_{X\ell} $.}
\label{fig:n2piXl} 
\end{figure}

{\bf An example neutron decay involving decuplet baryon $\Delta$}---%
For the purpose of illustration, we present a complete analysis of the final example decay $n\to\ell^-\pi^+ X$ in the Letter.
As shown in \cref{fig:n2piXl}, it
receives two contributions: a contact contribution due to the octet baryon Lagrangian in \cref{eq:chiB1} and a non-contact contribution due to the exchange of the $\Delta$ baryon from \cref{eq:chiT0}.
The contact contribution to the amplitude is,
\begin{align}
i\mathcal{M}_{(a)}=\frac{i 2 \sqrt{2} c_4}{F_0 \Lambda_\chi^2 \Lambda_{X\ell}^4} \big(\bar{u}_\ell \hat\Gamma^\tL_{\mu\nu\alpha\beta} u_n \big) p_X^\mu\, p_n^\alpha\, p_\pi^\beta\,
\epsilon_X^{\nu*},
\label{eq:Ma}
\end{align}
where $p_{n,\pi,X}$ are respectively the momentum of the neutron, pion, and dark photon and $\epsilon_X$ is the dark photon's polarization vector. 

The non-contact contribution due to the $\Delta^-$ pole is more delicate due to subtleties in describing a spin $3/2$ field in relativistic quantum field theory.
For definiteness, we adopt the so-called consistent chiral Lagrangian for the strong interaction among the octet and decuplet baryons and the octet pseudoscalar mesons~\cite{Geng:2009hh}:
\begin{align}
\mathcal{L}^{(1)}_{MBT}=\frac{i\mathcal{C}}{ M_D} \epsilon^{abc} D_\rho\bar{T}_\mu^{ade}\gamma^{\rho\mu\nu} (u_\nu)_{db} B_{ec}+\hc,
\end{align}
where $u^\mu=i\big(\xi(\partial^\mu-i r^\mu)\xi^\dagger-\xi^\dagger(\partial^\mu-il^\mu)\xi\big)$ and $\gamma^{\mu\nu\rho}=-(i/2)\{\sigma^{\mu\nu},\gamma^\rho\}$
\cite{MartinCamalich:2010nab}. The chiral covariant derivative of $T_\mu^{ade}$ is defined as $D_\nu T_\mu^{abd}=\partial_\nu T^{abc}_\mu + (\Gamma_\nu)_{ad} T_\mu^{dbc}+(\Gamma_\nu)_{bd} T_\mu^{adc}+ (\Gamma_\nu)_{cd} T_\mu^{abd}$.
The decuplet baryons are identified with the components of the totally symmetric tensor via  
\begin{align}
& T_{111} = \Delta^{++},~
T_{112} =\frac{ \Delta^{+}}{\sqrt{3}},~
T_{122} =\frac{ \Delta^{0}}{\sqrt{3}},~
T_{222} =\Delta^{-},~
\nonumber
\\
&T_{113} =\frac{\Sigma^{*+}}{\sqrt{3}},~
T_{123} =\frac{ \Sigma^{*0}}{\sqrt{6}},~
T_{223} =\frac{\Sigma^{*-}}{\sqrt{3}},~
T_{133} =\frac{\Xi^{*0}}{\sqrt{3}},~
T_{233} =\frac{\Xi^{*-}}{\sqrt{3}},~
T_{333} =\Omega^{-}.
\end{align}
The LEC $\mathcal{C}\approx1$ is determined by fitting the $\Delta\to N\pi$ decay width \cite{Pascalutsa:2006up}, and the decuplet-baryon mass $M_D$ is set to $1.382$ GeV. 
The interaction relevant to the black dot vertex in \cref{fig:n2piXl} (b) is,
\begin{align}
\mathcal{L}^{(1)}_{MBT}\supset \frac{i\sqrt{2}\mathcal{C}}{ M_D F_\phi} \epsilon^{abc} (\partial_\rho\bar{T})_\mu^{ade}\gamma^{\rho\mu\nu} (\partial_\nu\Pi)_{db} B_{ec},
\end{align}
where $F_\phi\equiv 1.17 \times 92.4~\rm MeV$ is an average of decay constants \cite{Geng:2009hh}. 
The propagator for the spin-3/2 Rarita-Schwinger field is \cite{Benmerrouche:1994uc}
\begin{align}
S_{\mu\nu}=-\frac{(\slashed{p}+m)}{p^2-m^2+i\epsilon}\left[ g_{\mu\nu}-\frac{1}{3}\gamma_\mu\gamma_\nu-\frac{1}{3m}(\gamma_\mu p_\nu-\gamma_\nu p_\mu) - \frac{2}{3m^2} p_\mu p_\nu \right].
\end{align}
Then the amplitude generated from the $\Delta^-$ baryon exchange is,
\begin{align}
i\mathcal{M}_{(b)}= \frac{2 \sqrt{2}i \tilde{c}_4 \mathcal{C}} { F_0 \Lambda_\chi M_D \Lambda_{X\ell}^4} \big(\bar{u}_\ell \hat\Gamma^\tL_{\mu\nu\alpha\beta} S^{\beta\eta} \gamma_{\rho\eta\delta}\, u_n \big) p_X^\mu\, q^\alpha\, q^\rho\, p_\pi^\delta\, \epsilon_X^{\nu*}.
\label{eq:Mb}
\end{align}
With \cref{eq:Ma,eq:Mb}, the decay width is
\begin{align}
\Gamma_{n\to\ell^-\pi^+ X}=\frac{1}{256\pi^3 m_n^3}  
\int_{s_-}^{s_+} d s\int_{t_-}^{t_+} d t ~\overline{|{\cal M}_{(a+b)}|^2},
\end{align}
where the upper and lower bounds of the integration variables $s=(p_\ell+p_X)^2$ and $t=(p_n-p_\ell)^2$ are 
\begin{align}
& 
s_-=(m_\ell +m_X)^2,\,\quad
s_+= (m_n - m_\pi)^2,
\notag\\
& t_\pm = (E_2^* + E_3^*)^2 - \Big(\sqrt{E_2^{*2} - m_X^2} \mp \sqrt{E_3^{*2} - m_\pi^2} \Big)^2, 
\notag\\
& E_2^* \equiv \frac{s - m_\ell^2 + m_X^2}{2\sqrt{s} },\quad
E_3^* \equiv \frac{ m_n^2 - s - m_\pi^2}{2\sqrt{s}}.
\end{align}
Finally, the numerical result is
\begin{align}
\Gamma_{n\to e^-\pi^+ X}= 
\big( 0.88|\kappa_4|^2 + 0.27 |\tilde \kappa_4|^2 - 0.82 \Re(\kappa_4 \tilde \kappa_4^*) \big)
\left(\frac{3\times 10^6\,\rm GeV}{\Lambda_{X\ell}}\right)^8 \frac{1}{10^{30}\,\rm yr},
\end{align} 
where $\kappa_4 \equiv c_4/c_1$ and
$\tilde\kappa_4 \equiv \tilde c_4/c_1$ parametrize the contributions from the octet baryon Lagrangian in \cref{eq:chiB1} and the decuplet baryon Lagrangian in \cref{eq:chiT0}, respectively. 
However, determination of all these new LECs is required to quantify their contributions.

\end{document}